# Optimal Intelligent Control for Wind Turbulence Rejection in WECS Using ANNs and Genetic Fuzzy Approach

[1] H. Kasiri, [2*] H. R. Momeni, [3] A. Kasiri

[1, 2] *Faculty of Electrical and Computer Engineering, Tarbiat Modares University, Tehran, Iran*

[3] *Department of computer science, Tehran University, Tehran, Iran*

Email: [1] hadi.kasiri@gmail.com, [2] momeni_h@modares.ac.ir, [3] Atiehk90@gmail.com

*Abstract.* One of the disadvantages in Connection of wind energy conversion systems (WECSs) to transmission networks is plentiful turbulence of wind speed. Therefore effects of this problem must be controlled. Nowadays, pitch-controlled WECSs are increasingly used for variable speed and pitch wind turbines. Megawatt class wind turbines generally turn at variable speed in wind farm. Thus turbine operation must be controlled in order to maximize the conversion efficiency below rated power and reduce loading on the drive-train. Due to random and non-linear nature of the wind turbulence and the ability of Multi-Layer Perceptron (MLP) and Radial Basis Function (RBF) Artificial Neural Networks (ANNs) in the modeling and control of this turbulence, in this study, widespread changes of wind have been perused using MLP and RBF artificial NNs. In addition in this study, a new genetic fuzzy system has been successfully applied to identify disturbance wind in turbine input. Thus output power has been regulated in optimal and nominal range by pitch angle regulation. Consequently, our proposed approaches have regulated output aerodynamic power and torque in the nominal range.

* Corresponding Author:
  H. R. Momeni,
Faculty of Electrical and Computer Engineering,
Tarbiat Modares University Jalal Ale Ahmad Highway, Tehran, Iran,
Email: mailto:momeni_h@modares.ac.ir

## 1. Introduction

One of the oldest energy conversion devices is wind turbine. These devices were the usual and fundamental energy conversion plants in past windmills. The mechanical energy required to mill has been achieved by these Wind Energy Conversion Systems (WECSs) successfully. At the present time, wind turbines are being established in many countries to produce electrical energy.

These instances reveal that people have used wind energy to provide essential services. Due to inflation, and facing the end of fossil fuels many countries hope to manufacture electric energy using natural and renewable resources such as geothermal, wind, sun and biomass caused by cheap and biological limitations.

Regarding to electricity production, one of the most hopeful renewable energy resources is wind energy [1].

Researchers estimate that the power producing of wind turbines will be increased further, especially in offshore and communal applications in the near future [2, 3].Consequently problems caused by the wind must be reduced using offered control approaches. In more cases, wind have turbulent that it creates vibration in wind turbine output, thus blades rotation notably does not raise from the upper bound of the rated speed. Similarly it does not decrease from the lower bound of the rated speed too. Therefore it is important to hold turbine power in a balanced range.

Whenever turbulence wind speeds takes place, the blade pitch control practically decreases the fluctuation of WECSs. There are several ways to determine suitable pitch angle for steady operation [13, 21]. In this paper, a suitable control signal is generated using the error between angles that produce a constant power output. There are different algorithms to control turbine output power. However, intelligent and online control is the best choice for identity and rejection of wind turbulence. A typical wind energy conversion system is capable of changing rotational speed and blade pitch angle is given as a block diagram in Figure. 1. In some cases this control is accomplished using generalized predictive models [4], while in other cases PID controllers are employed. In other word the benefits when operating in the immediate surroundings of gusty, nonlinear and adaptive model-based controllers, large wind speeds have been documented [6, 7].

Multivariable control describes an approach that afford pitch angle [8]. On the other hand, informed works exploit such adaptive controllers for efficient power conversion, related to maximum power point (MPP) tracking [9, 10], with less regard to the WECS's structural integrity. In [11], a feature-extraction algorithm, a frequency analyzer, was developed, and the features are formulated as the inputs of an artificial previous term neural network next term using back propagation. An artificial-previous term neural-network next term-based controller has been presented in [12] to realize fast valuing in a power-generation plant.

Progression of hybrid architectures [13], knowledge acquisition for symbolic AI systems and improved adequacy for data mining applications [14, 15], these show some purposes of rule extraction approaches are data exploration. Genetic algorithms (GAs) have been used in various problems, such as nonlinear optimization, combinatorial optimization and machine learning [16-18]. Also genetic algorithms are applied for selecting fuzzy if-then rules, modification of nonlinear scaling functions, and for determining hierarchical structures of fuzzy rule-based systems. A cascade GA [19], a micro-GA [20] is uncommon genetic algorithm that was used for designing fuzzy rule-based systems.

Due to random and non-linear nature of the wind turbulence and the ability of Multi-Layer Perceptron (MLP) and Radial Basis Function (RBF) Neural Networks (NNs) in the modeling and control of this turbulence, in this study, widespread changes of wind have been perused using MLP and RBF artificial NNs. In addition in this study, fuzzy rules have been successfully extracted from Neural Network (NN) using a new Genetic Fuzzy System (GFS). In addition a new genetic fuzzy system has been successfully applied to identify disturbance wind in turbine input. Thus output power has been regulated in optimal and nominal range by pitch angle regulation. Results indicate that the new proposed genetic fuzzy rule extraction system outperforms one of the best and earliest methods in controlling the output during wind fluctuation. Consequently these proposed approaches optimally have been adjusted turbine output power by pitch angle regulation. Simulated results of these two methods have verified that in comparison to other proposed approaches, our smart controller has reached its demands with higher accuracy in wind turbulence rejection.

In this study first we will declare components of WECSs modeling in section 2. MLP and RBF neural networks and Genetic Fuzzy Systems have been indicated in Section 4 and 5. Afterward our proposed methods in section 6 and 7 are completely asserted with simulation results. Finally we will compare these methods in section 8.

## 2. WECS modeling

*A. Characterization*

One of the input modules in turbine model is wind speed which rotates the blades. Full WECS model that consists of dynamic and electrical model has been be offered by [21]:

*1) Wind modeling or generated wind speed to time.*

In this part wind time pattern is formed founded on the ARMA series.

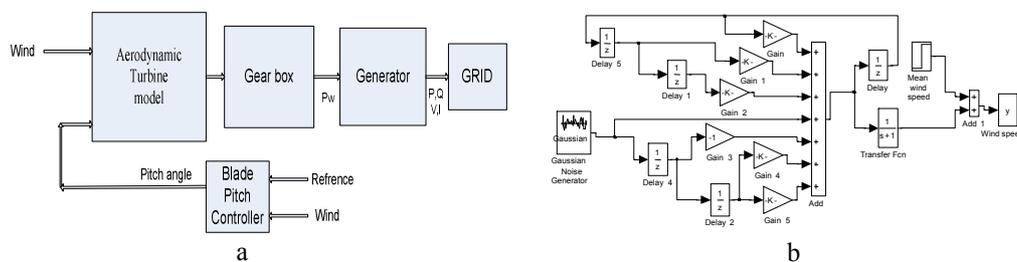

a            b

**Figure. 1.** (a) Block diagram of a WECS and (b) Generation of wind speed by the ARMA model.

**Table 1.** Specifications of the wind power generating facility

| Wind turbine and rotor | Pitch controller | Wind field |
|---|---|---|
| Blade radius, R 37.5  Number of blades  3  Cut-in/cut –out wind speed 4/25 m/s  Rated capacity,  2 MW | Max/min pitch angle 30/-2 deg  Max/min pitch rate 8/-8 deg/s | Rated wind speed  12 m/s  Air density  1.225 kg/$m^3$  Turbulence intensity  16% |

The wind speed $v_w(t)$ has two element parts declared as:

$$v_w(t) = v_m + v_t(t) \qquad (1)$$

where $v_m$ is the mean wind speed at hub height and $v_t(t)$ is the instantaneous turbulent part,

whose linear model is composed by a first-order filter disturbed by Gaussian noise. The instantaneous turbulence component of wind speed is obtained as [21]:

$$v_t(t) = \delta_t \vartheta_t \tag{2}$$

where $\delta_t$ is the standard deviation and $\vartheta_t$ is the ARMA time series model, which may be expressed as:

$$\vartheta_t = \delta_1 \vartheta_{t-1} + \delta_2 \vartheta_{t-2} + ... + \delta_n \vartheta_{t-n} + \eta_t - \theta_1 \eta_{t-1} - ... - \theta_m \eta_{t-m} \tag{3}$$

where $\eta_t$ is the white noise process with zero mean, $\delta_i (i=1,2,...,n)$ and $\theta_j (j=1,2,...,m)$ are the autoregressive parameters and moving average parameters, respectively. Wind speed by the ARMA model is generated in Simulink environment as Figure 1 (b).

*2) WECS control block*

This section contains mechanical controls for example blade pitch angle and power order to converter and electrical controls that include reactive power control and generator that connection with the network.

Table 1 reveals the simulation parameters, baseline turbine supposition, and safety operational limits of the WECS.

*B. Aerodynamic model*

The wind power available, the power curve of the machine and the ability of the machine to respond to wind fluctuation are the main factors in the power production of wind turbine. The presentment for power created by the wind is given by [22-24]:

$$P = \frac{1}{2}\rho \pi R^2 v_w^3 C_p(\lambda, \beta) \tag{4}$$

where $\rho$ is air density, $R$ is radius of rotor, $v_w$ is wind speed, $C_p$ denotes power coefficient of wind turbine, $\lambda$ is the tip-speed ratio and $\beta$ represents pitch angle. The relationship between performance $C_p$ with $\lambda$ and $\beta$ can be presented by [25]:

$$C_p(\lambda, \beta) = 0.5176 \left( \frac{116}{\lambda_i} - 0.4\beta - 5 \right) e^{-21/\lambda_i} + 0.0068\lambda \tag{5}$$

Note that $\lambda_i$ depends on instantaneous $\{\lambda, \beta\}$, hence tip-speed ratio and $\lambda_i$ are defined as:

$$\lambda = \frac{\omega_t R}{v_w} \quad \text{and} \quad \lambda_i = \frac{1}{\lambda + 0.8\beta} - \frac{0.035}{\beta^3 + 1} \tag{6}$$

where $\omega_t$ is the rotor speed. It can be seen that if the rotor speed is kept constant, then any changes in the wind speed will change the tip-speed ratio, therefore the changes of power coefficient $C_p$ controls as well as the manufactured power out of the wind turbine. If, the rotor speed is regulated according to the wind speed variation, then the tip-speed ratio can be preserved at an optimal point by pitch angle changes, which could produce maximum power output from the system.

In a control model the controlled torque $\Gamma_c$, which it has been used for the variable-speed turbines, is given by [26]:

$$\Gamma_c = K\omega^2 \tag{7}$$

where the gain K is given by

$$K = \frac{1}{2}\rho AR^3 \frac{C_{p\max}}{\lambda_*^3} \tag{8}$$

## 3. Artificial Neural Networks

The linear approximation in various widespread cases is not valid and the accuracy of system modeling decreases significantly. Thus ANNs are capable of modeling very complex functions. Thus Artificial Neural Networks (ANNs) are mathematical representations inspired by the functioning of the human brain [27-29]. In addition they keep in check the curse of dimensionality problem that bedevils efforts to model nonlinear functions with large numbers of variables [30].

## 4. MLP and RBF neural networks

Neural networks are very sophisticated modeling techniques, which are capable of modeling extremely complex functions. In particular, neural networks are nonlinear (a term which will be discussed in detail later in this section). For many years linear modeling techniques commonly have been used in most modeling domains and linear models have well-known optimization strategies. However in some frequent cases the linear approximation is not valid and the accuracy of system modeling decreases significantly. Neural networks also keep in check the curse of dimensionality problem that bedevils attempts to model nonlinear functions with large numbers of variables.

    *A. Multi-layer Perceptron networks*

The units each perform a biased weighted sum of their inputs and pass this activation level through a transfer function to produce their output, and the units are arranged in a layered feed forward topology. The network thus has a simple interpretation as a form of input-output model, with the weights and thresholds (biases) the free parameters of the model. Such networks can model functions of almost arbitrary complexity, with the number of layers, and the number of units in each layer, determining the function complexity. Important issues in MLP design include specification of the number of hidden layers and the number of units in these layers. The number of input and output units is defined by the problem. The number of hidden units to use is far from clear. As good a starting point as any is to use one hidden layer, with the number of units equal to half the sum of the number of input and output units. Training process and computation in layers and neurons happen by equation [31]:

$$y_p^{(k)} = sgm_p^{(k)}\left[W_{ip}^{(k-1)} \cdot y_i^{(k-1)} - \beta_i^k\right] \quad (p = 1,2,\ldots,N_k; k = 1,2,\ldots,M) \tag{9}$$

Where $W_{ip}^k$ is the connection weight between the ith neuron in the (k _ 1)th layer and pth neuron in the kth layer, $y_p$ the output of the pth neuron in the kth layer, $sgm_p$ the sigmoid activation function of the pth neuron in the kth layer and $\beta_p^k$ the threshold of the pth neuron in the kth layer. Sigmoid activation function is given as

$$sgm(x) = \frac{1}{1+\exp(-x)} \tag{10}$$

Training process of the back propagation algorithm runs according to the following steps [32, 33, 38, 39]:
1. Initialize all weights at random.
2. Calculate the output vector.
3. Calculate the error propagation terms.
4. Update the weights by using Eq. (11).
5. Calculate the total error '' $\varepsilon$ '' by using Eq. (13).
6. Iterate the calculation by returning to error is less than the desired error.

$$W_{ip}^{(k-1)}(t+1) = W_{ip}^{k-1}(t) + \alpha \sum_{n=1}^{I} \delta_{np}^{(k)} y_{ni}^{(k-1)} \tag{11}$$

where t is the iteration number and $\alpha$ is the learning rate and

$$\delta_{np}^{(k)} = sgm_{np}^k(.)\left[\sum_{n=1}^{I} \delta_{np}^{(k)} W_{pl}^{(k)}(t)\right] \tag{12}$$

$$\varepsilon = \sum_{n=1}^{I} \sum_{j=1}^{N_M} \left(y_{nj}^{(M)} - \hat{y}_{nj}^{(M)}\right)^2 \tag{13}$$

### B. Radial Basis Function networks

A radial basis function (RBF) network, therefore, has a hidden layer of radial units, each actually modeling a Gaussian response surface. Since these functions are nonlinear, it is not actually necessary to have more than one hidden layer to model any shape of function: sufficient radial units will always be enough to model any function. RBF networks have a number of advantages over MLPs. First, as previously stated, they can model any nonlinear function using a single hidden layer, which removes some design-decisions about numbers of layers. Second, the simple linear transformation in the output layer can be optimized fully using traditional linear modeling techniques, which are fast and do not suffer from problems such as local minima which plague MLP training techniques.

Radial basis Gaussian transfer function is considered as (14) in this study

$$F(u,c,\sigma) = \exp\left\{-\left(\frac{u-c}{\sigma}\right)^2\right\} \tag{14}$$

where c is the center, $\sigma$ is the variance and u is the input variable. The output of the ith neuron in the output layer at time n is

$$y_i = \sum_{j=1}^{H} W_{ij} F_j(u,c,\sigma)$$

(15)

Training process of the radial basis function neural network runs according to the following steps [32, 33].

1. Initialize all weights at random.
2. Calculate the output vector by Eq. (15).
3. Calculate the error term "$\varepsilon$" of each neuron in the outputlayer according to (16).

$$\varepsilon_i(n) = y_i(n) - \hat{y}_i(n); (i = 1,2,...,L) \tag{16}$$

where $\hat{y}_i$ is the desired output vector of the ith neuron in the output layer and $y_i$ the calculated output vector of the ith neuron in the output layer by using (14).

4. Update the weights by using Eq. (17).
5. Calculate the total error "$\varepsilon_T$" according to (18).
6. Iterate the calculation by returning to Step 2 until the total error is less than the desired error.

$$W_{ij}(n+1) = W_{ij}(n) + \alpha \varepsilon_i(n) F_j(u,c,\sigma); (i = 1,2,...,L; j = 1,2,...,H) \tag{17}$$

where n is the iteration and $\alpha$ the learning rate

$$\varepsilon_T = \sum_{n=1}^{I} \sum_{j=1}^{N_M} (y_{nj} - \hat{y}_{nj})^2 \tag{18}$$

## 5. Genetic Fuzzy Systems

Nowadays various problems, such as nonlinear optimization and machine learning have been solved by Gas [44]. Many GA-based methods have been planned for inbreeding fuzzy if-then rules.

Those methods can be classified into two approaches in the same manner as the categorization of non-fuzzy genetics-based machine learning methods: the Michigan approach and the Pittsburgh approach [34].

    1) Michigan approach

In the Michigan approach, only a single fuzzy rule-based system is stored during the execution of genetic algorithms, whereas a number of fuzzy rule-based systems are stored in the Pittsburgh approach [35]. This also leads to the short calculation time required for the generation update by genetic operations. Consequently one advantage of the Michigan approach is its small memory requisite. But one disadvantage of the Michigan approach is the deficiency of a straight coherency between the implementation of genetic algorithms and the optimization of fuzzy rule-based systems. Because the performance of each fuzzy rule-based population is not utilized in growing the fuzzy rule-based population, hence algorithm searched for suitable fuzzy if-then rules to perform its optimization indirectly.

    2) Pittsburgh approach

In the Pittsburgh approach, a set of if-then rules is coded as a string to which genetic operations such as selection, crossover, and mutation are applied [35]. One advantage of this approach is that the performance of each fuzzy rule-based system can be directly used as a fitness function. Because of a population consists of a number of fuzzy rule-based systems, Pittsburgh approach is its large memory necessity, thus its one disadvantage for this approach. The Pittsburgh approach requires much more memory storage than the Michigan approach, in which a single fuzzy-rule based system corresponds to the whole population.

## 6. Artificial Neural Networks controller strategy

NNs are mathematical representations inspired by the functioning of the human brain [48-49]. Without containing detailed analytical patterns for the systems, NNs could exert complex input–output mapping.

*A. Proposed strategy*

Variable speed is facilitated by pitch regulation to control the power extracted by the rotor. During high wind turbulence events, the extracted wind power has to be regulated via blade pitching.

Thus the pitch controller is activated when the wind speed exceeds or fall the rated speed, and the power is limited to its rated value since speed and power are coupled via the maximum power point tracking. In the proposed method we have used a predictive controller; because the learning process uses credible optimal values. Therefore these methods predict the value of pitch angle and take it to blades actuator. The goal is to train the neural network in such a way that desired output is produced. Our proposed approach rapidly gives the best output control signal to turbines in wind farms on different places. Thus it does not need to write new program and run it for these places. In other words our method does not depend on environment conditions. Since it is only related to initial data or learning data. This is the greatest advantage of our proposed method.

According to Figure. 2(a) the MLP controller in this paper has two inputs (wind speed and output power) and one output (pitch angle value). At first this controller is trained with credible optimal values of wind turbine input-output.

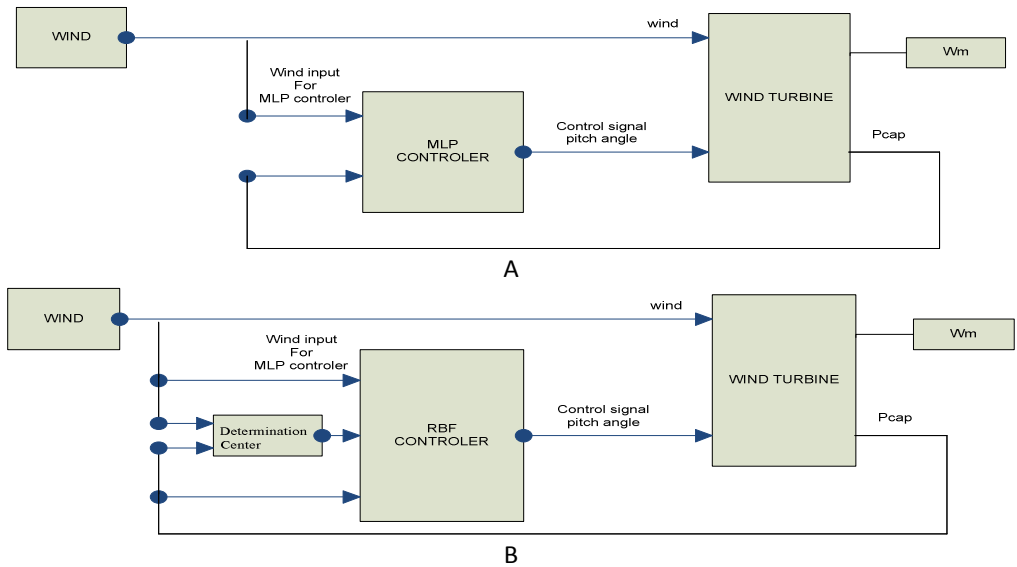

**Figure. 2.** Block diagram of controllers. (a) MLP controller diagram. (b) RBF controller diagram.

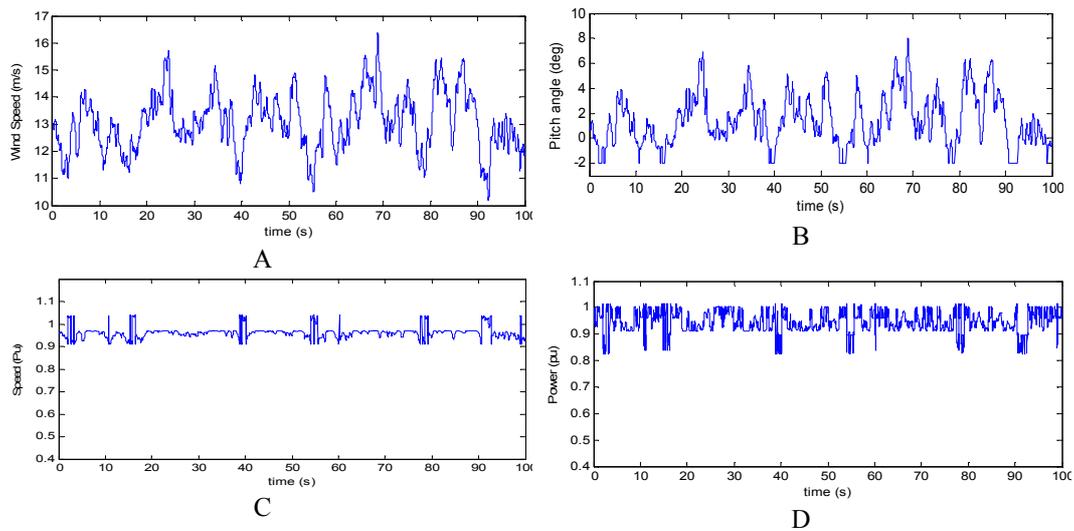

**Figure. 3.** Evolution of control and power alteration parameters in MLP Neural Network. (a) Simulated wind speed. (b) Pitch angle. (c) angular speed of wind turbine blades. (d) output power of wind turbine in Pu

Therefore this approach procures the best weight between input output using desired error produces. The training of NN computes an appropriate pitch angle upon catch wind speed. There are five neurons in the hidden layer. The hidden layer has nonlinear activation functions, but the output layer has a linear one.

The RBF neural network controller has three inputs and one output. Figure. 2(b) shows block diagram of RBF controller. Similar to previous controller at first this controller is trained with full credible rang optimal value of wind turbine input-output. Thus RBF controller results the best weight between input output using desired error produces, and produces an appropriate pitch angle upon catch wind speed. There are 10 neurons in the hidden layer. The hidden layer has nonlinear activation functions, but the output layer has a linear one.

### B. Simulation Results of Artificial Neural Networks controller

Simulation results brightly confirm the truth of the proposed control methods. In this section controller output and output power of wind turbine have been simulated. A wind time model is created founded on the ARMA series; the mean wind speed is determined based on spectral energy distribution of the wind and an overlaid noise signal. A wind speed shape and variation of it is painted in Figure. 3(a) and Figure. 4(a). Figure. 3 (b) reveals that increase in wind speed enhances in the blade pitch angle $\beta$. Whenever the wind speed decreases below the rated value of 12 m/s, $\beta$ is fixed in -2 degree. According to Figure. 3(c) the angular speed of wind has been controlled actually. As plotted nearby rated wind speed, at large wind speeds, a dynamic variation of the generator speed have been permitted by controller. Because of it absorbs rapid alterations in power through wind gusts thus escaping mechanical pressures. Afterward Figure. 3(d) presented output power of wind turbine in Pu, which has been managed by MLP controller perfectly.

As plotted power reduces in some spots that it has been resulting of intense and sudden wind speed reduction, but MLP algorithm with changes in pitch angle almost control it. Similarly, in RBF algorithm output simulation and analysis Figure. 4 (b) shows the blade pitch angle $\beta$ is increased by the pitch angle controller that generates the demanded command signal whenever the wind speed rises above the rated value of 12 m/s and inverse. Figure. 4 (c) reveals angular speed of wind turbine blades that has been controlled by RBF neural network. Lastly, Figure. 4 (d) presented output power of wind turbine in Pu, which has been regulated by RBF controller. Similar to MLP simulation output the power curve in RBF distinguishes power drops in some spots that it's result of turbulence wind speed. However, RBF algorithm has controlled it with change pitch angle. In addition, we observe respond of MLP NN shows a better result than RBF NN because the desired values in training have not cluster mode and variation of wind speed have the relatively large scatter.

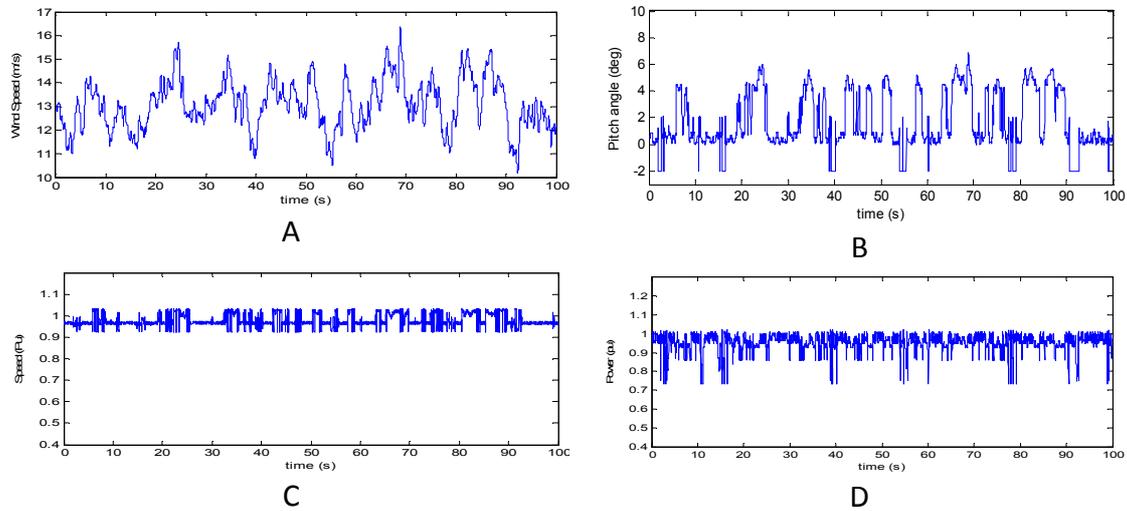

**Figure. 4.** Evolution of control and power alteration parameters in RBF Neural Network. (a) Simulated wind speed. (b) Pitch angle. (c) angular speed of wind turbine blades. (d) output power of wind turbine in Pu.

## 7. Proposed Genetic Fuzzy System (GFS) Strategy

In classifier systems to model classification problems, a fuzzy if-then rule or Strings Consist of Coefficient (SCCs) is handled as an individual in genetic algorithms. Thus we can view the classifier systems as a kind of Michigan approach.

*1) Classifier System*

Stride 1. Generate an initial population of fuzzy if-then rules or SCCs. Each of them is generated randomly. The resultant class and the certainty grade of each fuzzy if-then rule or SCC are exhibited by a heuristic rule discussed by Ishibuchi in [35].

Stride 2. Evaluate each fuzzy if-then rule or SCC in the current population.

Stride 3. Generate new fuzzy if-then rules or SCC by genetic operations such as selection, crossover, and mutation.

Stride 4. Replace a part of the current population with the newly generated fuzzy if-then rules or SCC.

Stride 5. If a pre-specified stopping condition is not satisfied, return to Stride 2.

*2) Coding of Fuzzy Rule*

In our proposed method fuzzy if-then rules are coded as a numeral strings. Since the consequent class and the certainty grade can be easily specified, each fuzzy if-then rule is implied by its random fuzzy sets.

The following symbols are used for denoting the seven linguistic values:
1: very small, 2: small, 3: medium small, 4: medium, 5: medium large, 6: large, 7: very large
For example, the following fuzzy if-then rule is coded as "17": If Xl is very small thenY1 is very large.

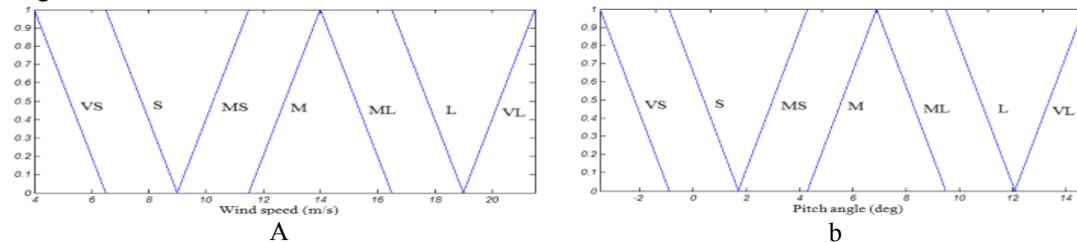

A    b

**Figure. 5.** Membership functions. (a) For wind speed. (b) For Pitch angle. VS, very small; S, small; MS, medium small; M, medium; ML, medium large; L, large; VL, very large.

*3) Definition of Fitness Function*

In fuzzy classifier systems, a population of fuzzy if-then rules corresponds to a fuzzy rule-based classification system. Fitness function in GFS includes two parts; one of them compares generated rules with optimal values, thus a rule that covers most of the optimal values could be a proper rule. In the other part; numeral equivalent of rules have calculated on wind turbine power formula, accordingly a rule that has the best control on power and regulates it well, could be a target rule. Finally each piece of these parts has been allocated a weight. As a result the best rule will be selected from initial random rules.

*4) Genetic Operations*

New fuzzy if-then rules have been generated from the current population by a pair of parent selected fuzzy if-then rules. The offspring of selected parents are generated by the uniform crossover. Hence each value of generated individuals is randomly replaced with other values by the mutation operator. A preferred number of fuzzy if-then rules are generated by iterating the selection, crossover, and mutation procedures. Finally the algorithm replaces the worst fuzzy if-then rules with the smallest fitness values with the newly generated fuzzy if-then rules that have the utmost fitness values.

**Table 2.** Project selection matrix rules

| *if* wind speed is | *Then* pitch angle is |
|---|---|
| MEDIUM SMAL | VERY SMALL |
| MEDIUM | SMALL |
| MEDIUM LARGE | MEDIUM |
| LARG | LARG |
| VERY LARG | VERY LARG |

The number of removed fuzzy if-then rules is usually the same as that of added rules. Finally,

GFS determines five rules for the control system. We illustrate these results in Table 2 and show membership functions in Figure. 5.

According to block diagrams of GFS controller in Figure. 6 controller has two inputs and one output. This controller provides a suitable pitch angle upon catch wind speed.

*5) Simulation Results*

To verify the effectiveness of the proposed control method, controller output and output power of wind turbine have been simulated. First wind speed has been simulated by MATLAB. Wind time series is generated based on the ARMA model. The mean wind speed is determined based on spectral energy distribution of the wind and a superimposed noise signal.

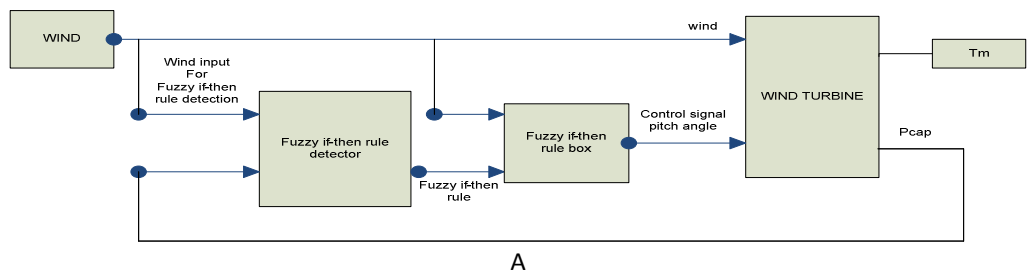

**Figure. 6.** Block diagram of controllers. (a) GFS controller diagram.

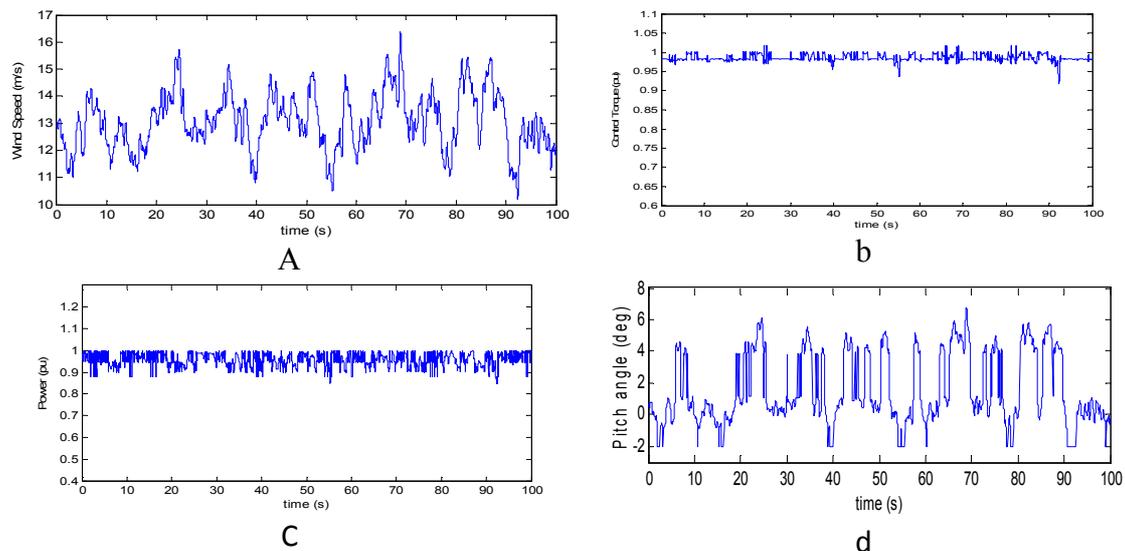

**Figure. 7.** Evolution of control and power alteration parameters in GFS and output power for two compared approaches. (a) Simulated wind speed. (b) Controlled torque. (c) output aerodynamic power of wind turbine in Pu. (d) Pitch angle signal for GFS.

A profile of the wind speed variation is depicted in Figure. 7(a). Figure. 7(b) depicts the established generator torque $\Gamma$ for GFS controller. The output mechanical power of GFS is shown in Figure. 7(c).

Figure. 7 (d) shows the blade pitch angle $\beta$ control signal for GFS controller. In below rated wind speed, optimal power is attained by regulating $C_{p-opt}$ (optimal performance coefficient); thus, the pitch angle is kept at a mechanical minimum and rotor speed is controlled in such a way that $\lambda_{opt}$ (optimal tip speed ratio) is always acquired, akin to MPP tracking.

## 8. Comparison

Proposed methods in this study have been trained to control turbine in wind fluctuation. Moreover output power has small lapse despite widespread wind mutations. In this study it has been tried to control turbine in wind turbulence. Consequently, output power has least variations. An output power leveling control strategy of wind farm based on both average wind farm output power and standard deviation of wind farm output power has been offered and used in the variable pitch angle wind turbine control system in [36]. Figureure 8(a), indicates output power of this strategy in 180 seconds. The System response has sudden change, and hence is sensitive to wind turbulence extremely. Simulation results of this method indicate that output power intensity had downfall and rise dramatically. The System response has sudden change, and decreases less than 0.7 Pu, hence according to Figure. 8 (e) and Figureure. 8 (f) our GFS method has improved output responds with genetic algorithm very well.

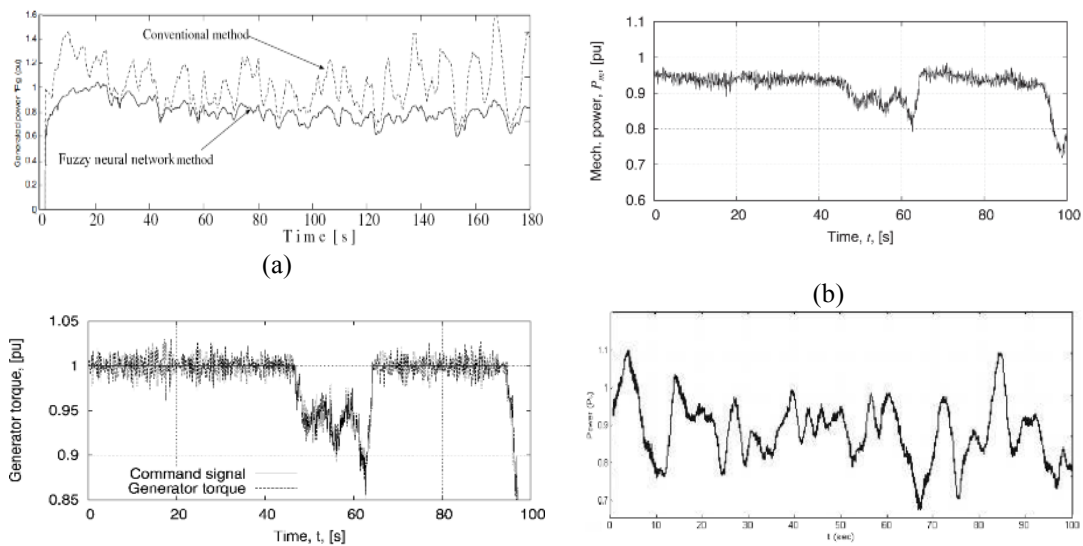

(a)

(b)

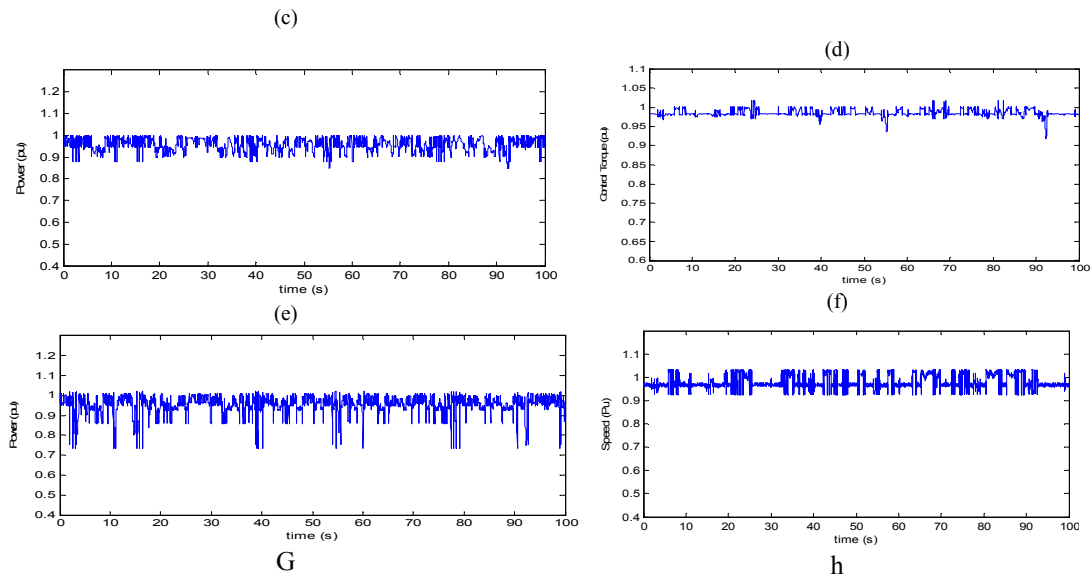

**Figure. 8.** Evolution of power alteration. (a) Output power for method based on the Fuzzy neural network in Pu. (b) Output power of LQG algorithm in Pu. (c) Generator torque of LQG algorithm in Pu. (d) output power for method based on the BP neural network (e) Output power of GFS in Pu. (f) Controlled torque of GFS in Pu. (g) output power of MLP controller in Pu. (h) controlled angular speed of wind turbine blade in MLP controller in Pu.

Classical methods based on PI(D) algorithms are a good starting point for many aspects of closed-loop controller design for variable-speed turbines. Control design is a task that demands rigorous test data and extensive engineering judgment. The presented approach in [21] has been employed to model the WECS subsystems as a basis for multi objective controller design. Model validation is carried out to authenticate the results. Produced power and generator torque from this strategy is shown in the Figure. 8 (b) and Figure. 8 (c). A PID controller based on the BP neural network has been offered and used in the variable pitch angle wind turbine control system in [9]. Figureure 8(d), indicates output power of this strategy. The System response has sudden change, and hence is sensitive to wind turbulence extremely. Simulation results of this method indicate that output power intensity had downfall and rise dramatically.

According to Figure. 8(f) the torque has been controlled actually. As plotted nearby rated wind speed, at large wind speeds, a dynamic variation of the generator speed have been permitted by controller. Because of it absorbs rapid alterations in power through wind gusts thus escaping mechanical pressures. Afterward Figure. 8(g) presented output power of wind turbine in Pu, which has been managed by MLP controller perfectly.

This method has saved power in same places, but it has severe change when power reduces in long time that is an effect of the wind downfall. Figure 8 (e) and Figure. 8 (f) shows that our proposed algorithms have produced controlled power and torque with fewer drops than output power and torque of LQG algorithm [21], for example, between 45 to 70 sec. In those situations that wind speed reduces intensely, output power downfall for a while. Note that this event is unavoidable. In the power curve, power reduction in some spots is distinguished that it's result of wind sudden reduction, but proposed algorithms with shift pitch angle almost control it.

Table 3. Comparision of two rule extraction algorithms

| Operation / Algorithms | Iteration | Mutation probability | Cross over probability | Population size | Membership size | Time required to Rule extraction | Lowest error in the Rule extraction |
|---|---|---|---|---|---|---|---|
| FRENGA | 50 | 0.7 | 0.4 | 65 | 7 | 302.646775 | 0.1226 |
| FRE without NN using GA | 100 | 0.6 | 0.7 | 60 | 7 | 501.212821 | 0.1445 |

Also we compare FRENGA [42, 43] by our GFS approach in Table 3. In FRENGA method we have used from neural network in fitness function. Despite the same simulation results and rules, FRENGA extracts rules with great speed but almost equal accuracy.

## 9. Conclusion

The controllability of wind power generation is necessary for the combination of wind energy into the power system. Because of one of the world's fastest increasing source of clean and renewable energy is wind energy. In this article, two new methods estimate and predict pitch angle value for a wind turbine introduced. These models have one input: wind speed. Hence their suitability in wind turbulences controlling is investigated. Proposed controllers are trained using multi-layer perceptron and radial basis function neural networks. As plotted from simulation results, controllers change pitch angle of turbine blades to achieve most suitable output power.

According to comparison results, MLP Neural Network Controller has better respond than RBF Neural Network and both better than BP neural network. Proposed controllers successfully trained with full credible and optimal rang value of wind turbine input-output. Consequently an optimal weight between input and output is produced.

MLP Neural Network Controller and GFS Controller have been compared to some other learning algorithms using difference between the power outputs. As it is demonstrated GFS Controller has the better respond. Hence, the required limitation during wind turbulences was provided by the proposed controllers. These algorithms trained with full credible rang optimal value of wind turbine input-output successfully. Finally controllers optimally produced accurate rules or curve between input and output. Thus it is clear that output power of wind turbine does not have sudden and frequent changes and capable to connect to central network with less effect of disturbance on it. For this method a limitation is that we need a computer with high speed and high memory for record previous data.

In this paper, blade pitch position control has been submitted by two smart controllers during wind turbulence. Practically, both the approaches yielded acceptable results. Consequently, considering that the desired values in training, have not cluster mode and have relatively large scatter, MLP-based controller donated better results than the other type.

## 10. Future work

We must tuned membership function to improve result and accuracy. This work will cause that speed process increase.

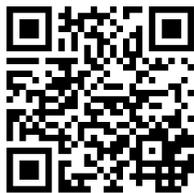

Free download this article and more information